# Material combination optimization for brazed ceramic-metal composites using Artificial Intelligence

Sunita Khod[1], Vinay Kamma[2], Ravi Kumar Verma[3], and Mayank Goswami[1, 2, #]

[1]*Divyadrishti Imaging Laboratory, Department of Physics,* Indian Institute of Technology Roorkee, Roorkee, Uttarakhand-247667, India.

[2]*Mehta Family School of Data Science and Artificial Intelligence*, Indian Institute of Technology Roorkee, Roorkee, Uttarakhand-247667, India.

[3]*Process & Materials Quality Assurance Division–Mechanical (PMQD-M)*, Space Applications Centre, Indian Space Research Organization, Ahmedabad-380015, India

[#]mayank.goswami@ph.iitr.ac.in

## Abstract

This study proposes an Artificial Intelligence (AI) driven methodology for predicting a combination of brazed ceramic-metal composite materials. Multiple machine learning (ML) algorithms are compared with the deep learning (DL) model. The developed models are tested using k-fold validation. Nine different input-output feature configurations are evaluated to assess the model performance. The input-output feature comprises material properties, namely, the coefficient of thermal expansion (CTE) and molecular mass of brazed ceramic-metal composite materials obtained from literature and the strength parameter (average Von Mises Stress (VMS)) estimated from Finite Element Method (FEM) simulation for joint assembly structure. A multi-output model, Autoencoder (AE), has also been developed and tested to predict various features.

The ML model, namely the polynomial regression (PR), outperforms the other ML/DL models with a Mean square Error (MSE) of 0.01 for the test data. The autoencoder model with a 32-16-32 structure outperforms LR, PR, RF, and ANN with an MSE of 0.04% for the prediction of unseen data. The developed multi-output model accurately predicts all the features (single and multiple), while PR fails to accurately predict multi-output features of low importance. The developed AE model predicts the different material properties with an average error of ~0.16-3.78%  with literature-reported values.

Keywords: Finite Element Method, composite materials, Machine Learning, and Artificial Intelligence.

## 1. Introduction

Ceramic-metal composite materials are used in the aerospace and space industry to create components for satellites, radars, and spacecraft [1], [2]. One of the challenges is to develop long-lasting joints made of several components, each made either of alloys, pure metals and/or ceramics, glass, and/or polymers/ceramics. The components of such a composite joint may not have similar material properties. For example, dissimilar CTE of ceramic and metal materials may harbor residual stress at the joint interface region. The accumulation of such residue during extreme cyclic mechanical and thermal stress may result in early failure[3].

The brazing technique, one of the candidate fabrication processes, employs an apt filler material to balance mismatch in *certain* material properties[4]. This filler material can be a metal or metal alloy with a melting point lower than the base materials [5], [6]. The process requires heating ceramic, filler, and metal components in a high-temperature environment. Subsequently, cooling down to room temperature. The filler material fills the gap between the base materials through capillary action during the heating. The solidification of the filler material during the cooling process produces ceramic-metal joints [7].

## 1.1. Material property optimization

The overall performance of the joint can be improved by selecting the *optimal* combination of base and filler materials. The combination that can bear higher thermal stress and hence possess lower *effective* CTE is defined as optimal in this study[8], [9]. The usual method is to fabricate the brazed ceramic-metal composite joint assembly of effective CTE value (represented by $\alpha_{eff}$ ) and *assess* the strength/quality. The process is repeated for different material combinations until satisfactory statistical confidence for optimal strength/quality is established. The whole process is time-consuming and costly.

The quality assessment method can either be Destructive Testing (DT), Non-destructive Testing (NDT), or simulation or their combination [10], [11]. The DT uses conventional tests such as tensile, compression, or impact testing in a laboratory and estimates a specimen's shear, ultimate, or tensile strength. However, DT is time-consuming, destroys several copies of the specimen, generates waste, and requires high-cost investments. The NDT methods can estimate internal defects that can be correlated with the material properties/strength.

A simulation model correlated with real-world experiments can also create data [9]. Designers may estimate material properties/strength such as Von Mises stress and/or strain from simulation. Alternatively, the previously obtained data (optimality criteria and material properties) can be curve-fit to develop an analytical equation. The classical optimization method can be used to interpolate/extrapolate by minimizing/maximizing the optimality criteria to predict the combination.

## 1.2. Role of AI

Above all steps require human intervention to decide the operating parameters of experiments, thresholds for calculations, etc. For example, various stages of the strength/quality assessment of brazed ceramic-metal composite material joint assembly where a person is involved in making a choice may include: (a) selection of apt combination of base and filler material, (b) selection of fabrication technique of components, (c) strength assessment for the first decision, and (d) iteratively change '(a)' and (b), if '(c)' results in low strength. However, the number of iterations may be reduced with the experience of the expert. As the shape of the joint changes, the standard database created by the above methodology (section 1.2) and expert experience may not work.

An AI model can be used to augment and preserve the human experience. The AI model can be trained with the data obtained from steps (a) and (c). These AI models learn and remember the behavior of the brazed ceramic-metal composite materials and their corresponding strength/quality assessment parameters from the training data. The trained AI model can *predict* the appropriate combination of brazed ceramic-metal composite materials for a given strength value, thereby removing the need for iterative fabrication and strength/quality assessment. The AI may automate the process, fastening the process, saving material wastage, and reducing manual error.

### 1.3. Material property optimization, prediction, and AI

AI models are categorized into Machine Learning (ML) and Deep Learning (DL) models. The various ML models include Linear Regression (LR), Polynomial Regression (PR), Random Forest (RF), Support Vector Machine (SVM), Gradient Boosting, K-nearest neighbors (KNN), etc. At the same time, Artificial Neural Networks (ANN), Convolutional Neural Network (CNN), Recurrent Neural Network (RNN), Graph Neural Network (GNN), etc. are the types of DL models [12], [13]. ML models perform well for small data sets, while DL-based neural networks handle complex, high-dimensional data [14]. An Autoencoder (AE) is also a type of ANN and belongs to the class of DL models.

ML-based LR is used for structured and numeric type data for the prediction of continuous variables, while RF, SVM, KNN, GB, etc., are used for structured and tabular data for classification. DL-based neural networks such as U-net, RNN, and several other variant models work well for feature extraction and image segmentation [15]. ANN is a supervised learning algorithm that captures the non-linear and complex relationships in a training data set of numeric type in tabular form, making it superior to other ML and DL models [16]. An autoencoder (AE) utilizes a multi-layer neural network architecture, enabling the model to capture complex, non-linear patterns in the data through unsupervised training.

The data collected from DT by cameras during testing or NDT data in images requires convolutional neural networks such as U-net for segmentation and image processing. The ML-based models, such as LR, PR, RF, SVM, GB, etc., and DL-based models, such as ANN and AE, classify the DT and NDT data as numeric values to the class as per the training. An AE model can be trained using DT, NDT, or a simulation dataset with numerical input variables. Unlike conventional ANN models that directly predict output variables, autoencoders learn to compress the input data into a latent representation (encoding) and then reconstruct it (decoding), enabling the model to capture the most relevant features of the data. Once trained, the encoder part of the model can be used to extract reduced feature sets, detect anomalies, or serve as a pre-processing step for other machine learning tasks. Depending on the application, autoencoders can also be optimized using various techniques such as gradient descent, particle swarm optimization, or evolutionary algorithms.

The data for selecting optimal brazed ceramic-metal composite material may be obtained from DT, NDT, or simulation, and classical or AI methods can be used for optimization and selection. Therefore, depending on the data and method used, the process is classified as (a). DT assessed classically driven, (b). DT assessed AI-driven, (c). NDT assessed classically driven, (d). NDT assessed AI-driven, (e). simulation assessed classically driven, (f). simulation assessed AI-driven.

The classical driven ((a), (c), and (e)) optimization methods optimize the material properties by minimizing/maximizing the optimality criteria (tensile strength for DT, Porosity for NDT, or stress for simulation). The AI-driven ((b), (d), and (f)) optimization methods optimize the output parameter (material properties) during training and predict the accurate output for given input parameters during testing. Qi Zhang et al. reported the mechanical characterization of brazed $Al_2O_3$ joints fabricated with a Ni–Ti interlayer using experiment and simulation [18]. The characterization methods and improving the reliability of brazed ceramic-metal joint assemblies are explained in detail in a study by Ruixiang et al. [19]. Several studies for classical driven methods using DT [20], [21], NDT [22], [23], and simulation [24] are also reported for brazed ceramic-metal components and assemblies.

Sukhomay Pal et al. developed an ANN to predict the joint strength of weld joints fabricated using a pulsed metal inert gas welding process [25]. Mingoo Cho et al. modeled an ANN to predict the tensile strength of friction stir welded dissimilar materials [26]. A study to predict the yield stress and ultimate tensile strength of an aluminum alloy is also reported [27]. The use of neural networks utilizing data accessed from various methods for material property prediction of single material, alloys, or composite materials is reported by multiple research groups [28], [29], [30], [31], [32]. However, research studies utilizing AI-driven methods/neural networks for predicting the material properties of single or multi-material used for fabricating brazed ceramic-metal composite material joint structures for aerospace and higher temperature applications have not yet been reported.

### 1.4. Motivation

The process of selecting brazed ceramic-metal composite materials for optimized strength of joint assemblies needs to be automated. A feedback loop may be incorporated using the strength/quality assessed data as a function of brazed ceramic-metal composite material properties to predict the new material or combination of materials. This study proposes an AI-driven methodology for predicting the brazed ceramic-metal composite materials for fabricating joint assemblies for high-temperature and space applications (e.g., TWTA).

## 2. Material and Methods

The flowchart for the proposed method is shown in Fig.1. The process is divided into five steps. The first step arranges and annotates the material property data of brazed ceramic-metal composite materials for various types with their strength/quality parameters of the joint assembly in hand. The second step involves the essential feature extraction from the data set and SHAP (Shapley Additive exPlanations) value analysis. The next step involves modeling the different ML/DL models: LR, PR, RF, and ANN. An autoencoder (AE), which is a DL-based multi-output model, has also been developed. All the ML/DL models, including AE, are trained and tested in the fourth step. In the end, the performance of all the models for testing data is evaluated by computing Mean Square Error (MSE), and the best model's prediction is compared with the developed universal model (AE). Finally, the best model is used to predict the material properties of brazed ceramic-metal composite materials for given input feature values.

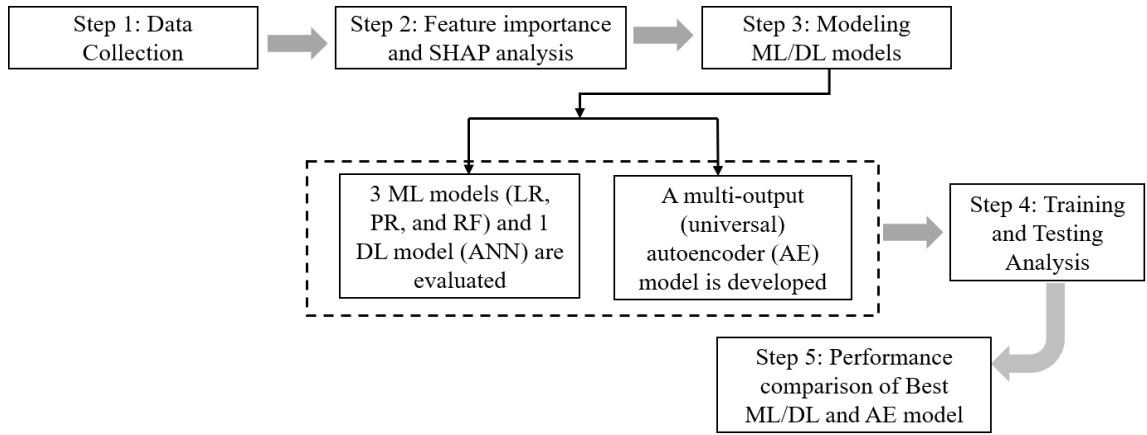

**Fig. 1.** Flowchart of the proposed methodology.

The proposed AI-driven methodology is implemented using libraries such as pandas, torch, sklearn, NumPy, etc., in Python. The pseudo-code for implementing the proposed AI models is illustrated in Algorithm 1. The essential libraries are initially imported, and the data set is loaded. After loading the data, the essential features are analyzed using feature importance and SHAP value analysis. Next, the nine different configurations (N1-N9) of input-output features are defined depending on the feature variables in the dataset. The input-output features include the microstructure property (Porosity) of the joint assembly, strength/quality parameter (average Von Mises Stress ($VMS$)), effective CTE ($\alpha_{eff}$) of the composite materials and material properties of one or more materials (CTE and molecular mass).

After the initial set-up, the four AI (ML/DL) models (LR, PR, RF, and ANN) and one autoencoder model are implemented. The ML/DL models are trained and tested using k-fold validation, and evaluation matrices are computed for nine configurations. The AE model is trained using a split dataset with 80 % training and 20 % validation. The trained AE model is then used to predict and plot the values for masked features. The prediction performance of the best ML/DL model is compared with the AE (universal) model.

> **Algorithm 1**: *AI modeling for ceramic-metal composite material property prediction*
> **Input**: micro-structure properties (porosity), strength parameter (VMS), material property of one or two material ($\alpha_{eff}$ and $Z$)
> **Ouput**: Material property of one or more material
> **Step 1: Library import and load dataset**
> - *import libraries*
> - *Load dataset* with input and output features
>
> **Step 2: Feature importance and SHAP value analysis**
> - Plot *feature importance* and *SHAP value* plot
> - *drop irrelevant* features
> - *arrange columns* keeping important features
>
> **Step 3: Define nine configurations for input-output features**
> Create a dictionary '*N_configs*':
>   Each key *(N1 to N9)* maps to a tuple:
>   - List of input features
>   - List of output targets
>
> **Step 4: Define ML/DL and multi-output Autoencoder (AE) model**
> *Define models:*
> - **LR, PR with degree 3, RF, and ANN**
> - Initialize '*saved_models*' dictionary to store trained models
> - Initialize result storage dictionaries: *results_r2, results_mse, results_rmse, results_mae*
>
> *Define AutoEncoder class:*
> - **Encoder: Linear → ReLU → Linear**
> - **Decoder: Linear → ReLU → Linear**
> - **Forward pass: Encode → Decode → Return output**
>
> **Step 5: Train, test and evaluate each model for each configuration**
> For each configuration *N_name* in *N_configs*:
>   - Get input and output variables (X, y)
>   - Create *k-fold* cross-validator (**5 splits**)
>   - Fit the model using **Multioutput Regressor**
>   - Predict on test data
>   - Evaluate metrices: $R^2$ and MSE
>   - Plot and save in *result dictionaries*
>   - *Save* the trained model in '*saved_models*'
>
> For each configuration of AE model:
>   - *Create* augmented dataset with *MaskedDataset*
>   - Split dataset: **80%** train, **20%** validation
>   - *Initialize DataLoaders* for training and validation
>   - Train and evaluate model with *early stopping*
>   - Predict for each *masked feature*
>   - *Plot* actual vs predicted value for each *masked feature*
>
> **Step 6: Compare the feature prediction plots of best ML/DL model with AE (universal) model**

## 2.1. Data collection

A complex high voltage feedthrough assembly of Travelling Wave Tube Amplifier (TWTA) and one simplest geometry combination of cubes are used to create brazed ceramic-metal composite joint assemblies. Different materials, including $Al_2O_3$, Kovar, Monel, Copper, aluminum, steel, and alloys are included. The material property data of candidate brazed ceramic-metal composite materials is obtained from the literature [33], [34], [35], [36]. The Finite Element Method simulation estimates thermal stress due to heating the joint structure/component/assembly from 300K to 1200K. The average Von Mises Stress (*VMS*) parameter calculated from the simulation is used as a decisive parameter for the strength/quality assessment of components fabricated from brazed ceramic-metal composite materials [9]. The *VMS* is estimated for different brazed ceramic-metal composite material combinations with joint structure/component/assembly containing various microstructure properties (pore volume, size, location, and Porosity). Fig. 2 (a1) and Fig. 2 (b1) show the joint assembly structures. Figures 2 (a2) and 2 b(2) depict the boundary conditions. Figs. 2 (a3) and 2(b3) show the simulated stress distribution for joint assembly containing alumina ($Al_2O_3$) ceramic, Ag-Cu-Ti braze alloy, and Kovar metal in the brazed ceramic-metal composite joint assembly. Fig. 2 (a4) and (b4) show the real-world assemblies manufactured by industries for various applications and whose structures resemble those we have incorporated in our study.

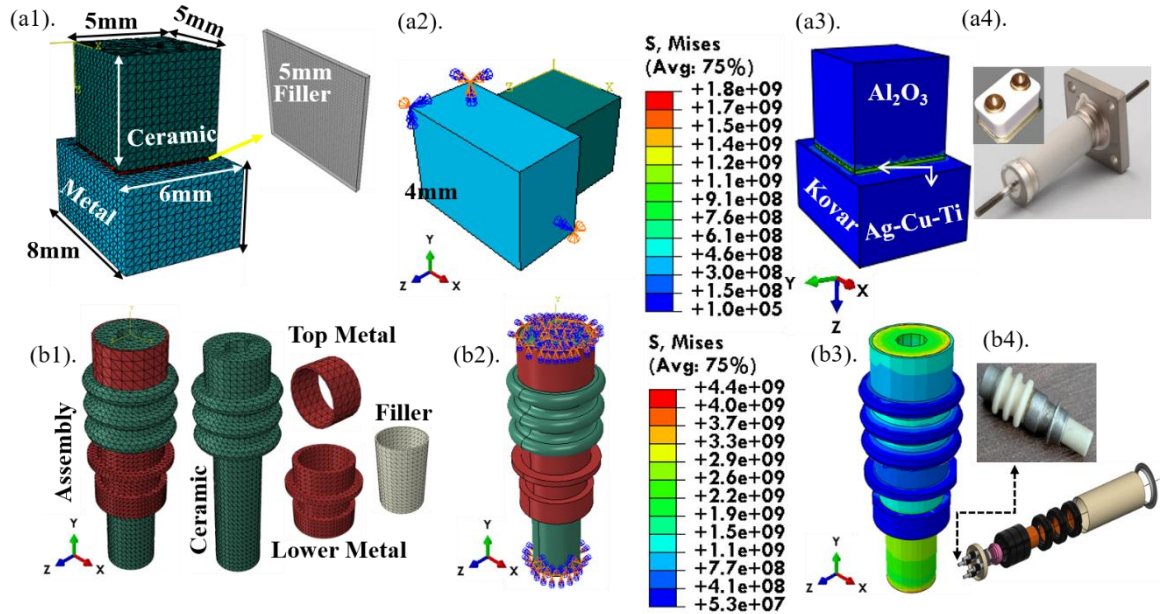

**Fig. 2.** Information of the joint assembly design and materials: (a1). Structure of the joint assembly, (a2). Boundary conditions for the assembly are shown in (a1), and (a3). Simulated stress distribution using FEM (a4). Real-world assembly with a structure similar to the designed one shown in (a)1 [37], [38], (b1). The joint assembly used in TWTA, (b2), boundary condition for joint assembly shown in (b1), (b3). Simulated stress distribution, and (b4). Joint assembly component similar to (b1) is used in TWTA [39].

The simulation data set contains the parameters pore volume, size, location, Porosity, $VMS$, effective CTE ($α_{eff}$), effective molecular mass ($Z_{eff}$), molecular mass of ceramic, braze, metal (represented $Z_C$, $Z_B$, $Z_M$), CTE of ceramic, braze, and metal (represented by $α_C$, $α_B$, $α_M$). Our recently published work gives a detailed description of obtaining material property and simulation data [9]. The real data set containing 88 points is obtained from the simulation. However, the material properties are sampled around two distinct values, creating a sparse variation and formation of a discrete cluster. Thus, a high overfitting was observed during early experiments with model training. Therefore, we introduced synthetic data points obtained from the interpolation method, and the final data contains 500 data points. The augmented data contains the feature variability and ensures smoother gradients. The principal component analysis (PCA) and t-distributed Stochastic Neighbor Embedding (t-SNE) plots are plotted and shown in Fig. SF0 of supplementary section S0. The PCA plot shows the global similarity and variation, while the t-SNE plot depicts the preservation of local neighborhood in synthetic data. The high feature importance and SHAP value parameters are considered when developing the AI models.

## 2.2. Modeling method

Next, the structure of the used AI model is defined after the data set preparation. Depending upon the combination of material property and feature importance, the input-output feature configuration of the model may vary. Therefore, we have defined nine configurations, namely N1, N2,..., and N9, of the input-output feature variables for the four ML/DL models. The details of the nine configurations are given in Table 1.

**Table 1:** The input-output feature combinations for different configurations of ML/DL models.

| S.No. | Configuration | No. of inputs | Input features | No. of outputs | Output features |
|---|---|---|---|---|---|
| 1. | N1 | 6 | Porosity, $VMS$, $α_{eff}$, $Z_{eff}$, $Z_C$, $Z_B$ | 1 | $Z_M$ |
| 2. | N2 | 6 | Porosity, $VMS$, $α_{eff}$, $Z_{eff}$, $Z_B$, $Z_M$ | 1 | $Z_C$ |
| 3. | N3 | 6 | Porosity, $VMS$, $α_{eff}$, $Z_{eff}$, $Z_C$, $Z_M$ | 1 | $Z_B$ |
| 4. | N4 | 4 | Porosity, $VMS$, $α_{eff}$, $Z_{eff}$ | 3 | $Z_C, Z_B, Z_M$ |
| 5. | N5 | 6 | Porosity, $VMS$, $α_{eff}$, $Z_{eff}$, $α_C$, $α_B$ | 1 | $α_M$ |
| 6. | N6 | 6 | Porosity, $VMS$, $α_{eff}$, $Z_{eff}$, $α_B$, $α_M$ | 1 | $α_C$ |
| 7. | N7 | 6 | Porosity, $VMS$, $α_{eff}$, $Z_{eff}$, $α_C$, $α_M$ | 1 | $α_B$ |
| 8. | N8 | 4 | Porosity, $VMS$, $α_{eff}$, $Z_{eff}$ | 3 | $α_C, α_B, α_M$ |
| 9. | N9 | 9 | Porosity, $α_{eff}$, $Z_{eff}$, $Z_C$, $Z_B$, $Z_M$, $α_C$, $α_B$, $α_M$ | 1 | $VMS$ |

The correlation between the material property and VMS is a regression problem for modeling brazed ceramic-metal joint assembly. The frequently used ML regression models include Linear Regression (LR), Polynomial Regression (PR), Support Vector Regression (SVR), Gradient Boosting Regression (GBR), Random Forest (RF), etc. The DL bases Artificial Neural Network (ANN) have shown significant performance for prediction over small data sets containing complex variables. Accordingly, we have developed and evaluated three ML models, namely LR, PR, and RF, and one DL model, namely ANN, as a part of our study. The details of four ML/DL models are given in Table 2.

**Table 2:** Details of ML/DL models.

| S.No. | ML/DL model | Parameters |
|---|---|---|
| 1. | LR | fit intercept=True |
| 2. | PR | degree=3, interaction_only=False, include_bias=True |
| 3. | RF | n_estimators =100, random_state = 42, min_samples_split = 2, min_samples_leaf=1 |
| 4. | ANN | hidden_layer_sizes = (16, 16), max_iter = 1000, early_stopping = False, validation_fraction = 0.1, n_iter_no_change = 20, learning_rate_init = 0.0001, random_state = 42 |

The ML/DL models must be trained and tested separately for each configuration of input-output features listed in Table 1. However, Autoencoder allows to predict the single and multiple output variables simultaneously by masking the essential features and is, therefore, an efficient method. Hence, we also developed a multi-output Autoencoder (AE) model to predict the features (material properties in our study) according to the requirement. The AE model is designed to learn efficient latent representations of material property and VMS data characterized by 10 distinct parameters, each representing a specific property or feature. The information on the input features and masked features for the developed AE model is given in Table 3.

**Table 3:** Details of input and masked features for nine configurations of the AE model.

| S. No. | Configuration | No. of inputs | Input features | No. of masked features | Masked feature (Output feature) |
|---|---|---|---|---|---|
| 1. | N1 | 9 | Porosity, $VMS$, $α_{eff}$, $α_C$, $α_B$, $α_M$, $Z_{eff}$, $Z_C$, $Z_B$ | 1 | $Z_M$ |
| 2. | N2 | 9 | Porosity, $VMS$, $α_{eff}$, $α_C$, $α_B$, $α_M$, $Z_{eff}$, $Z_B$, $Z_M$ | 1 | $Z_C$ |
| 3. | N3 | 9 | Porosity, $VMS$, $α_{eff}$, $α_C$, $α_B$, $α_M$ $Z_{eff}$, $Z_C$, $Z_M$ | 1 | $Z_B$ |
| 4. | N4 | 7 | Porosity, $VMS$, $α_{eff}$, $α_C$, $α_B$, $α_M$, $Z_{eff}$ | 3 | $Z_C, Z_B, Z_M$ |
| 5. | N5 | 9 | Porosity, $VMS$, $α_{eff}$, $Z_{eff}$, $Z_C$, $Z_B$, $Z_M$, $α_C$, $α_B$ | 1 | $α_M$ |
| 6. | N6 | 9 | Porosity, $VMS$, $α_{eff}$, $Z_{eff}$, $Z_C$, $Z_B$, $Z_M$, $α_B$, $α_M$ | 1 | $α_C$ |
| 7. | N7 | 9 | Porosity, $VMS$, $α_{eff}$, $Z_{eff}$, $Z_C$, $Z_B$, $Z_M$, $α_C$, $α_M$ | 1 | $α_B$ |
| 8. | N8 | 7 | Porosity, $VMS$, $α_{eff}$, $Z_{eff}$, $Z_C$, $Z_B$, $Z_M$ | 3 | $α_C, α_B, α_M$ |
| 9. | N9 | 9 | Porosity, $α_{eff}$, $Z_{eff}$, $Z_C$, $Z_B$, $Z_M$, $α_C$, $α_B$, $α_M$ | 1 | $VMS$ |

The N1-N8 input-output feature configuration of developed models predicts the features that are the material property of brazed ceramic-metal composite materials as a function of joint assembly strength. The N9 feature configuration is the reverse of N1-N8 configurations and predicts the strength parameter, which is $VMS$ of the joint assembly as a material property function.

The encoder component of the model receives the 10-dimensional input vectors and processes them through two fully connected linear layers. The first linear layer reduces the dimensionality of input data to predefined hidden units, followed by the ReLU activation function. The second layer further compresses the data to half the hidden units to produce a latent representation. The decoder then reconstructs the input data from the compressed latent representation to the original input dimension using two connected linear layers with a ReLU activation in between. The architecture of the multi-output AE model is given in Fig. 3. To optimize the model, three hidden layer configurations or hyperparameter combinations, namely [8, 4, 8], [16, 8, 16], and [32, 16, 32] are tested. The Adam optimizer with a learning rate of 0.001 is used as the optimization algorithm in the training. The number of epochs used in training is 250 with early stopping criteria and patience of 30 epochs, and the minimum delta threshold is $10^{-5}$.

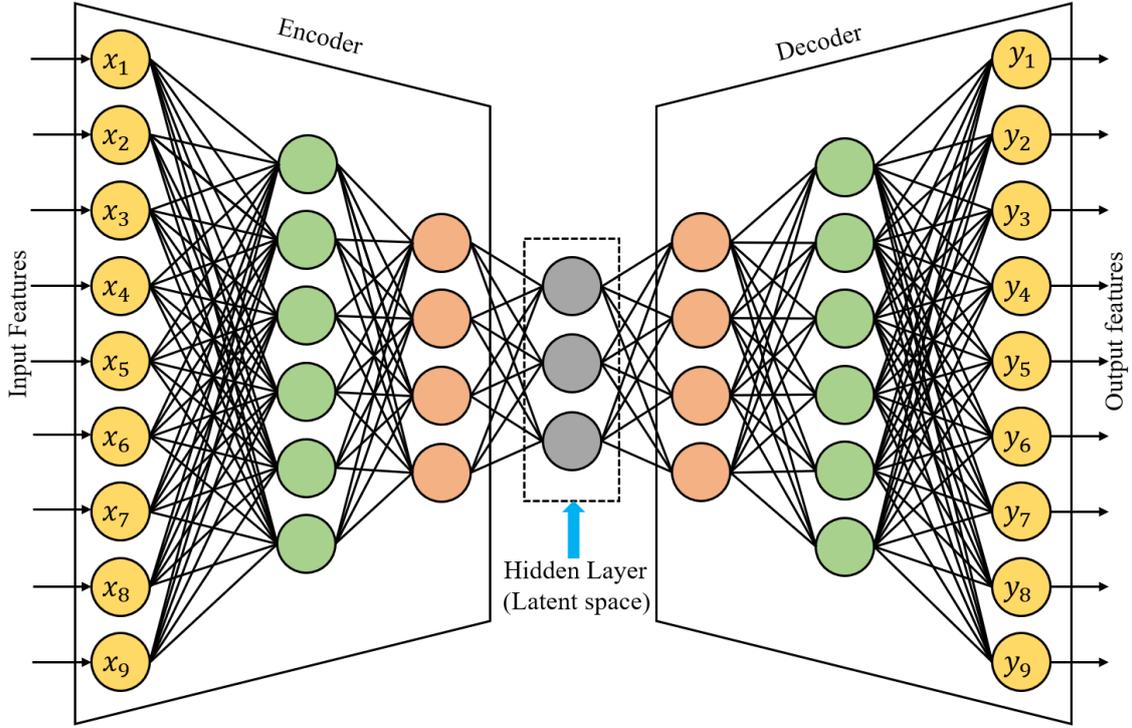

**Fig. 3.** Architecture of multi-output Autoencoder (AE) model.

### 2.3. Training and Testing Analysis

Finally, the training and testing analysis are evaluated after the data set preparation and modeling of the network structure of the ML/DL/AE models. The three ML models (LR, PR, RF) are trained and tested using k-fold validation with five splits and random state 42, while ANN is implemented with train and test splits of 90 % and 10 %, respectively. The AE model is trained with 80% of the data and tested with 20%.

The coefficient of determination ($R^2$), Mean Square Error ($MSE$), Root Mean Square Error ($RMSE$), and Mean Absolute Error ($MAE$) are used as evaluation metrics to forecast the performance of the different AI models. The value of $R^2$ depicts how well a trained model fits the data set. $MSE$ measures an average of the squared differences between predicted and actual values, while $RMSE$ is the square root of $MSE$. $MAE$ evaluates the average of the absolute differences between predicted and actual values. A model with high $R^2$ (~ 0.9-1) and low $MSE$, $RMSE$, and $MAE$ values on the training and test sets indicate that it can perform accurate predictions. The $R^2$, $MSE$, $RMSE$, and $MAE$ parameters are calculated as follows [40]:

$$R^2 = 1 - \frac{\sum_{i=1}^{n}(y_i^{act} - y_i^{pred})^2}{\sum_{i=1}^{n}(y_i^{act} - y_{avg})^2} \quad (1a)$$

$$MSE = \frac{1}{n}\sum_{i=1}^{n}(y_i^{act} - y_i^{pred})^2 \quad (1b)$$

$$RMSE = \sqrt{\frac{1}{n}\sum_{i=1}^{n}(y_i^{act} - y_i^{pred})^2} \quad (1c)$$

$$MAE = \frac{1}{n}\sum_{i=1}^{n}|y_i^{act} - y_i^{pred}| \quad (1d)$$

$$y_{avg} = \frac{1}{n}\sum_{i=1}^{n} y_i^{act} \qquad (1e)$$

Where $y_i^{act}$ is the value of material property reported in the literature and $y_i^{pred}$ is the material property predicted from the trained model. $y_{avg}$ is the average value of the material property parameter, and $n$ is the number of points in the data set.

The evaluation matrices are computed for all the ML/DL and AE models. The prediction accuracy of the best ML/DL model is compared with the developed optimized AE model.

## 3. Results

### 3.1. Feature Importance and SHAP value analysis

The material property influences the strength/quality (VMS) of the brazed ceramic-metal composite material joint assembly. Therefore, to analyze the contribution of each parameter or feature, feature importance, and SHAP value analysis is performed. The feature importance and SHAP value plot are shown in Fig. 4. The feature importance plot in Fig. 4(a) highlights that the molecular mass of metal ($Z_M$) material and effective molecular mass ($Z_{eff}$) of the joint assembly are the most influential parameters, followed by $\alpha_M$, Porosity, $Z_C$, $\alpha_C$, $\alpha_{eff}$, $Z_B$, and $\alpha_B$. The SHAP value plot in Fig. 4(b) represents the magnitude and direction of a parameter/feature's impact. It is observed that the braze material property, namely, $\alpha_B$ and $Z_B$ contribute more in positive directions while $Z_M$ and $Z_{eff}$ exhibit bidirectional impact.

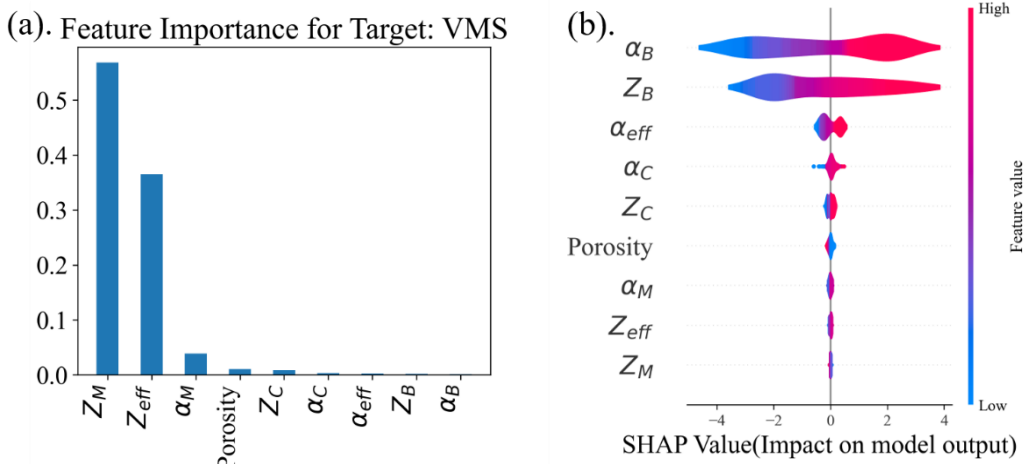

**Fig. 4.** Feature importance plot (**a**) and SHAP value analysis (**b**).

The feature importance parameter represents the global importance of a parameter identifying which features contribute more to the overall prediction of a model. The SHAP value plot represents the combined effect of global as well as local importance of a feature. These plots depict that the metal and braze material are essential materials for the performance of a brazed ceramic-metal composite material joint assembly.

## 3.2. ML/DL model Evaluation

The performance of the developed models (LR, PR, RF, and ANN) for predicting features, mainly the material properties of different materials, is evaluated. The evaluation metrics $R^2$, $MSE$, $RMSE$, and $MAE$ are calculated for all four ML/DL models and the input-output feature configurations (N1-N9). The bar plot for comparison of $R^2$ and $MSE$ values of LR, PR, RF, and ANN for N1 configuration are given in Fig. 5. The plot for $RMSE$, and $MAE$ is given in Fig. SF1 of supplementary files section S1.

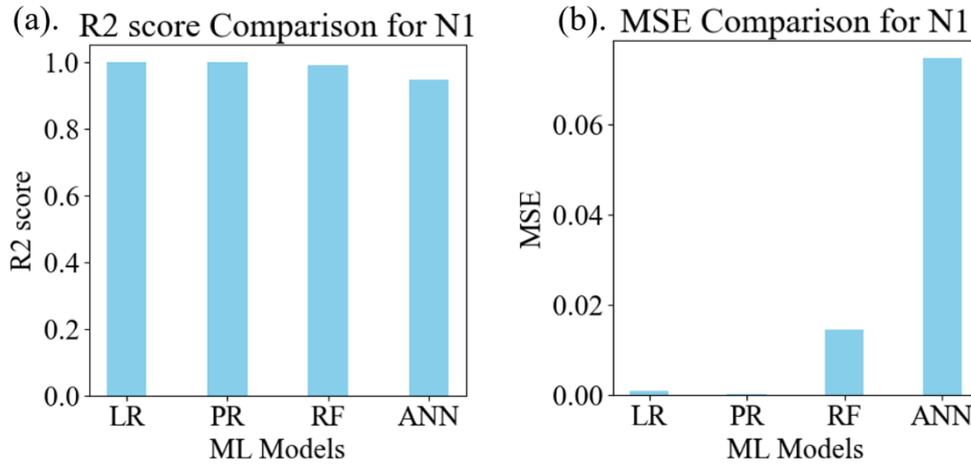

**Fig. 5.** Comparison of evaluation metrics for LR, PR, RF, and ANN for N1 configuration: (**a**). $R^2$ and (**b**). $MSE$.

The $R^2$ comparison plot in Fig. 5(a) shows that all the four models have $R^2$ values close to 1, with LR and PR having the highest values. The $MSE$ comparison plot in Fig. 5(b) depicts that the LR and PR have very low $MSE$ values, with PR achieving the lowest. The RF model has a slightly higher $MSE$ values, while ANN achieves the highest $MSE$. That means that the PR outperforms all the four ML/DL followed by LR as it achieves high $R^2$ and low $MSE$, and ANN shows the highest prediction errors (largest $MSE$ values). Hence, the PR model may be preferred for predicting features with individual outputs.

## 3.3. Optimization of AE model

The performance of the multi-output model known as AE for predicting the material property is also evaluated. To optimize the model, three hyperparameters are tested. The evaluation metrics $R^2$, $MSE$, $RMSE$, and $MAE$ are calculated and plotted. The plot for $R^2$ and $MSE$ values for different hyper-parameters are shown in Fig. 6, while the plot for $RMSE$, and $MAE$ values are given in Fig. SF2 of supplementary file section S2. It is observed that the [32, 16, 32] hidden layer configuration has the highest $R^2$ (~0.95) value and lowest $MSE$ (~0.04) values. Therefore, this hyperparameter configuration ([32, 16, 32]) is the optimal structure of the AE model and is used to predict multi-output variables. This optimal structure is referred to as AE_32.

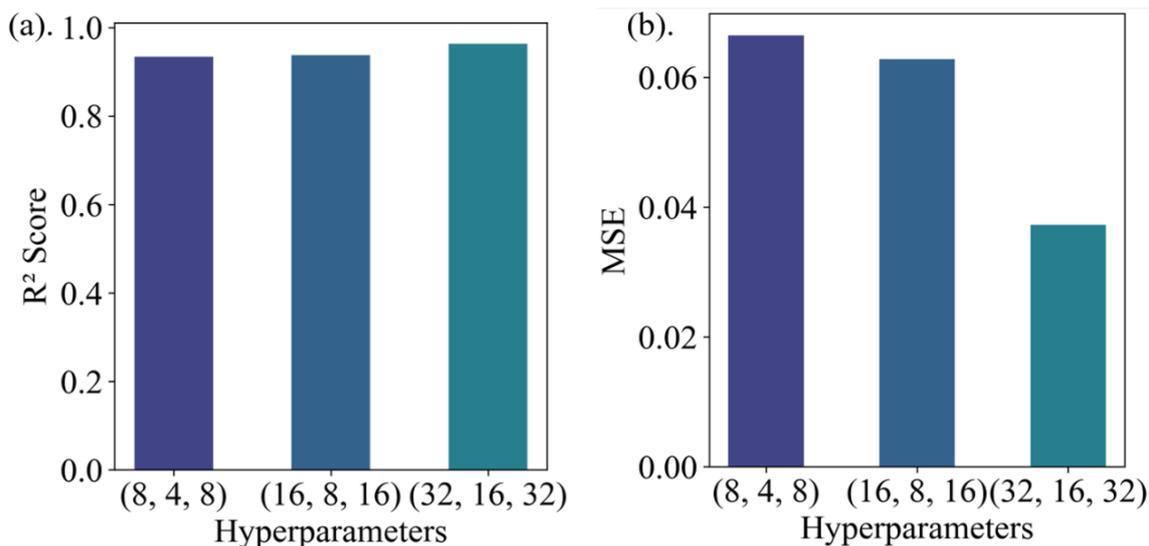

**Fig. 6.** Evaluation metrices comparison for different hyperparameters of AE model: (**a**). $R^2$ and (**b**). $MSE$.

After the hyperparameter optimization for the AE model, the number of epochs is varied up to 250 epochs with early stopping, and loss is calculated for the AE_32 model. It is observed that the loss ($MSE$) value decreases to ~0.1 and, saturates after 25 epochs and stops at 75 epochs. The loss vs epoch for AE_32 is plotted and shown in Fig. 7.

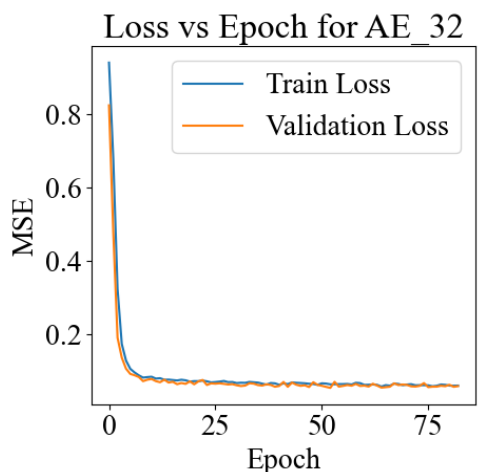

**Fig. 7.** Loss vs Epoch for AE_32 ([32, 16, 32] hidden layer configuration) model.

### 3.4. Prediction from best ML/DL vs. AE model

The evaluation metrices comparison for the ML/DL model suggests that the PR model best predicts the material properties for individual feature configurations. The AE model, which is a unified multi-output, also has a high prediction performance for various features. Therefore, the prediction from the implemented PR and AE models is plotted and compared. The plot for the prediction from the AE model and PR model for N1 configuration, which consists of the molecular mass of metal ($Z_M$) as output feature or variable is plotted and shown in Fig. 8. Fig. 8(a) shows the prediction from the AE model, while Fig. 8(b) contains the prediction plot from the PR model for $Z_M$.

It is observed that the best-fit equation for prediction from AE and PR models is $y = 0.92x + 4.75$ and $y = 1.00x + (-0.11)$, respectively. The predicted value lies around the best-fit equation for both models, which depicts their accuracy.

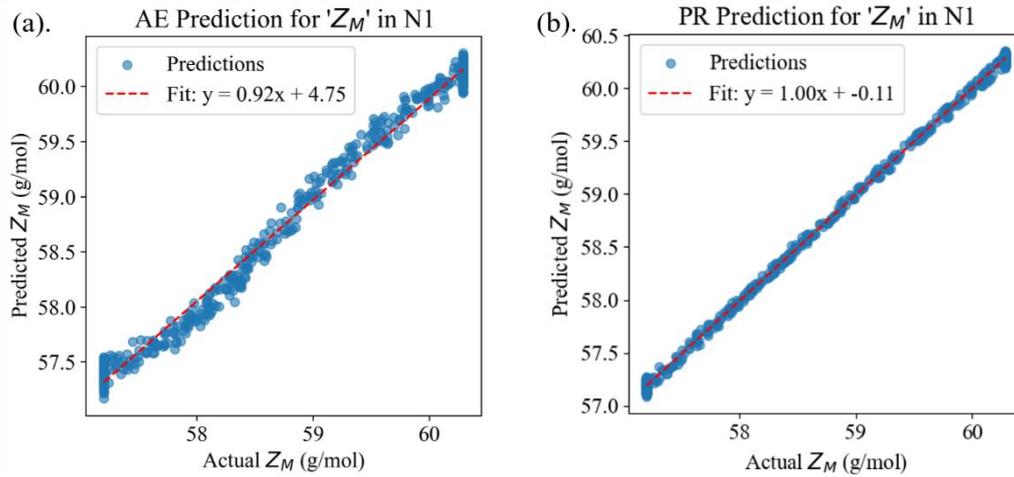

**Fig. 8.** Comparison between the predicted and actual values for $Z_M$ (molecular mass of metal): (**a**). AE model, (**b**). PR model.

The predicted value of the molecular mass of metal ($Z_M$) from the developed AE model for some given random input features is shown in Table 4, while the $Z_M$ values from the PR model are given in Table 5.

**Table 4:** Predicted $Z_M$ for random input parameters/features from the AE model.

| Input features | | | | | | | | | Masked feature | | Error (%) |
|---|---|---|---|---|---|---|---|---|---|---|---|
| Porosity (%) | VMS (MPa) | $\alpha_{eff}$ (°K$^{-1}$) | $\alpha_C$ (°K$^{-1}$) | $\alpha_B$ (°K$^{-1}$) | $\alpha_M$ (°K$^{-1}$) | $Z_{eff}$ (g/mol) | $Z_C$ (g/mol) | $Z_B$ (g/mol) | Act. $Z_M$ (g/mol) | Pred. $Z_M$ (g/mol) | |
| 0.25096 | 149 | 1.1×10$^{-5}$ | .770×10$^{-5}$ | 1.98×10$^{-5}$ | 1.53×10$^{-5}$ | 70.14 | 104.41 | 84.16 | 59.97 | 59.87 | 0.17 |
| 0.161211 | 151 | .79×10$^{-5}$ | .328×10$^{-5}$ | 1.94×10$^{-5}$ | 1.06×10$^{-5}$ | 74.68 | 140.31 | 86.9 | 58.27 | 58.34 | 0.12 |

**Table 5:** Predicted $Z_M$ for random input parameters/features from the PR model.

| Input parameters/features | | | | | | Output parameter | | Error (%) |
|---|---|---|---|---|---|---|---|---|
| Porosity (%) | VMS (MPa) | $\alpha_{eff}$ (°K$^{-1}$) | $Z_{eff}$ (g/mol) | $Z_C$ (g/mol) | $Z_B$ (g/mol) | Act. $Z_M$ (g/mol) | Pred. $Z_M$ (g/mol) | |
| 0.149551 | 158 | 9.94×10$^{-6}$ | 71.97 | 117.62 | 86.86 | 59.57 | 59.56 | 0.02 |
| 0.025186 | 113 | 9.76×10$^{-6}$ | 69.13 | 102.96 | 89.96 | 58.42 | 58.39 | 0.05 |

The feature variable for N2 configuration of models that is $Z_C$ which is referred to as the molecular mass of the ceramic material component of brazed ceramic-metal composite material joint assembly, is also predicted by AE and PR models. The plot for the actual vs. predicted $Z_C$ values from AE and PR models are plotted and given in

Fig. 9. The plot highlights that the best-fit curve fits the prediction points from AE and PR models with equations $y = 0.91x + 10.70$ and $y = 1.00x + (0.11)$ respectively highlighting the validity of the developed models.

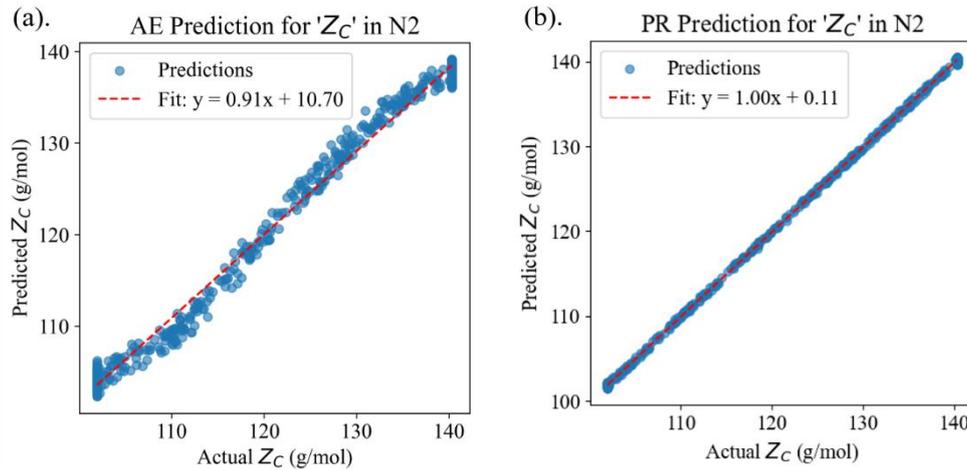

**Fig. 9.** Comparison between the predicted and actual values for $Z_C$ (molecular mass of ceramic material): (**a**). AE model, (**b**). PR model.

The predicted value of the molecular mass of ceramic material ($Z_C$) from the developed AE and PR models for some random input parameters, which are listed in Table 6 and Table 7, respectively.

**Table 6:** Predicted $Z_C$ for random input parameters/features from the AE model.

| Input features | | | | | | | | | Masked feature | | Error (%) |
|---|---|---|---|---|---|---|---|---|---|---|---|
| Porosity (%) | VMS (MPa) | $\alpha_{eff}$ (°K$^{-1}$) | $\alpha_C$ (°K$^{-1}$) | $\alpha_B$ (°K$^{-1}$) | $\alpha_M$ (°K$^{-1}$) | $Z_{eff}$ (g/mol) | $Z_B$ (g/mol) | $Z_M$ (g/mol) | Act. $Z_C$ (g/mol) | Pred. $Z_C$ (g/mol) | |
| 0.258981 | 200 | .968×10$^{-5}$ | .389×10$^{-5}$ | 2.01×10$^{-5}$ | 1.62×10$^{-5}$ | 74.93 | 81.79 | 60.29 | 135.27 | 135.73 | 0.34 |
| 0.283245 | 152 | .947×10$^{-5}$ | .584×10$^{-5}$ | 2.01×10$^{-5}$ | 1.26×10$^{-5}$ | 71.77 | 81.79 | 59.01 | 119.52 | 119.30 | 0.184 |

**Table 7:** Predicted $Z_C$ for random input parameters/features from the PR model.

| Input parameters/features | | | | | | Output parameter | | Error (%) |
|---|---|---|---|---|---|---|---|---|
| Porosity (%) | VMS (MPa) | $\alpha_{eff}$ (°K$^{-1}$) | $Z_{eff}$ (g/mol) | $Z_B$ (g/mol) | $Z_M$ (g/mol) | Actual $Z_C$ (g/mol) | Predicted $Z_C$ (g/mol) | |
| 0.311 | 181.0 | .845×10$^{-5}$ | 74.99 | 95.91 | 59.01 | 137.73 | 137.58 | 0.109 |
| 0.283245 | 152 | .947×10$^{-5}$ | 71.77 | 81.79 | 59.01 | 119.51 | 119.47 | 0.033 |

The molecular mass of braze alloy material ($Z_B$) for N3 input-output feature configuration is predicted from the AE and PR model. The plot for the comparison is shown in Fig. 10. It is observed that the best-fit equation for the prediction from the AE and PR model is $y = 0.81x + 16.88$ and $y = 0.99x + 0.19$ respectively. The slight deviation of prediction for $Z_B$ feature from both the models is due to the lower feature importance of the parameter.

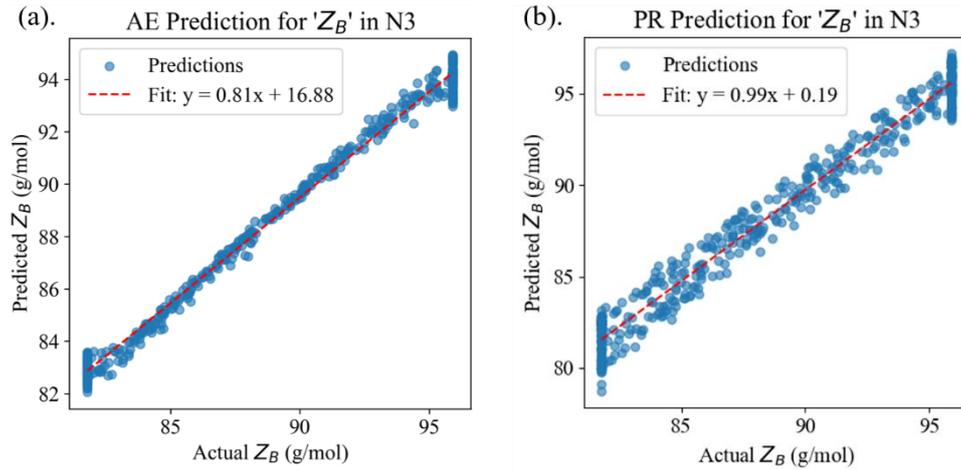

**Fig. 10.** Comparison between the predicted and actual values for $Z_B$ (molecular mass of braze material): (**a**). AE model, (**b**). PR model.

The predicted value of the Molecular mass of braze ($Z_B$) for some random input parameters/features from the developed AE and PR models, which are given in Table 8 and Table 9.

**Table 8:** Predicted $Z_B$ for random input parameters/features from the AE model.

| Input features | | | | | | | | | Masked feature | | Error (%) |
|---|---|---|---|---|---|---|---|---|---|---|---|
| Porosity (%) | VMS (MPa) | $\alpha_{eff}$ (°K⁻¹) | $\alpha_C$ (°K⁻¹) | $\alpha_B$ (°K⁻¹) | $\alpha_M$ (°K⁻¹) | $Z_{eff}$ (g/mol) | $Z_C$ (g/mol) | $Z_M$ (g/mol) | Act. $Z_B$ (g/mol) | Pred. $Z_B$ (g/mol) | |
| 0.223191 | 149 | 1.13×10⁻⁵ | .773×10⁻⁵ | 2.01×10⁻⁵ | 1.62×10⁻⁵ | 70.23 | 104.11 | 60.29 | 81.79 | 82.53 | 0.90 |
| 0.248857 | 142 | .767×10⁻⁵ | .328×10⁻⁵ | 1.96×10⁻⁵ | .961×10⁻⁵ | 74.34 | 140.31 | 57.92 | 85.12 | 85.97 | 0.99 |

**Table 9:** Predicted $Z_B$ for random input parameters/features from the PR model.

| Input parameters/features | | | | | | Output parameter | | Error (%) |
|---|---|---|---|---|---|---|---|---|
| Porosity (%) | VMS (MPa) | $\alpha_{eff}$ (°K⁻¹) | $Z_{eff}$ (g/mol) | $Z_C$ (g/mol) | $Z_M$ (g/mol) | Actual $Z_B$ (g/mol) | Pred. $Z_B$ (g/mol) | |
| 0.373416 | 132.0 | 9.77×10⁻⁶ | 70.52 | 110.37 | 58.87 | 89.44 | 89.14 | 0.34 |
| 0.047411 | 130.0 | 7.36×10⁻⁶ | 74.0 | 140.31 | 57.46 | 82.98 | 82.10 | 1.07 |

The N4 input-output configuration is a multi-output configuration that is the output features/variables in this configuration includes $Z_M$, $Z_C$, and $Z_B$ respectively. The developed AE and PR model are used to predict the $Z_M$

, $Z_C$, and $Z_B$ features. The comparison plot between the actual values and predicted values of these parameters from AE and PR model is plotted and shown in Fig. 11.

The best fit curve equation for the predictions of $Z_M$, $Z_C$, and $Z_B$ from AE model are $y = 0.88x + 7.07$, $y = 0.87x + 16.06$, $y = 0.77x + 19.91$ respectively while $y = 1.00x + 0.04$, $y = 1.00x + 0.53$, $y = 0.32x + 59.96$ are the best fit equations for prediction from PR model. The comparison plot in Fig. 11 highlights that the prediction from AE model are accurate for all the three parameters/features with a slight deviation from the linear fit. However, the PR model predicts well for $Z_M$ and $Z_C$ parameters with high feature importance but fails to predict for $Z_B$ parameter as shown in Fig. 11 (b3). The inaccurate predictions from PR for multi-output configurations while accurate predictions from AE model suggests that the AE model is good fit for both single output and multi-output configurations while developed PR model fits well for prediction of single output parameters/features.

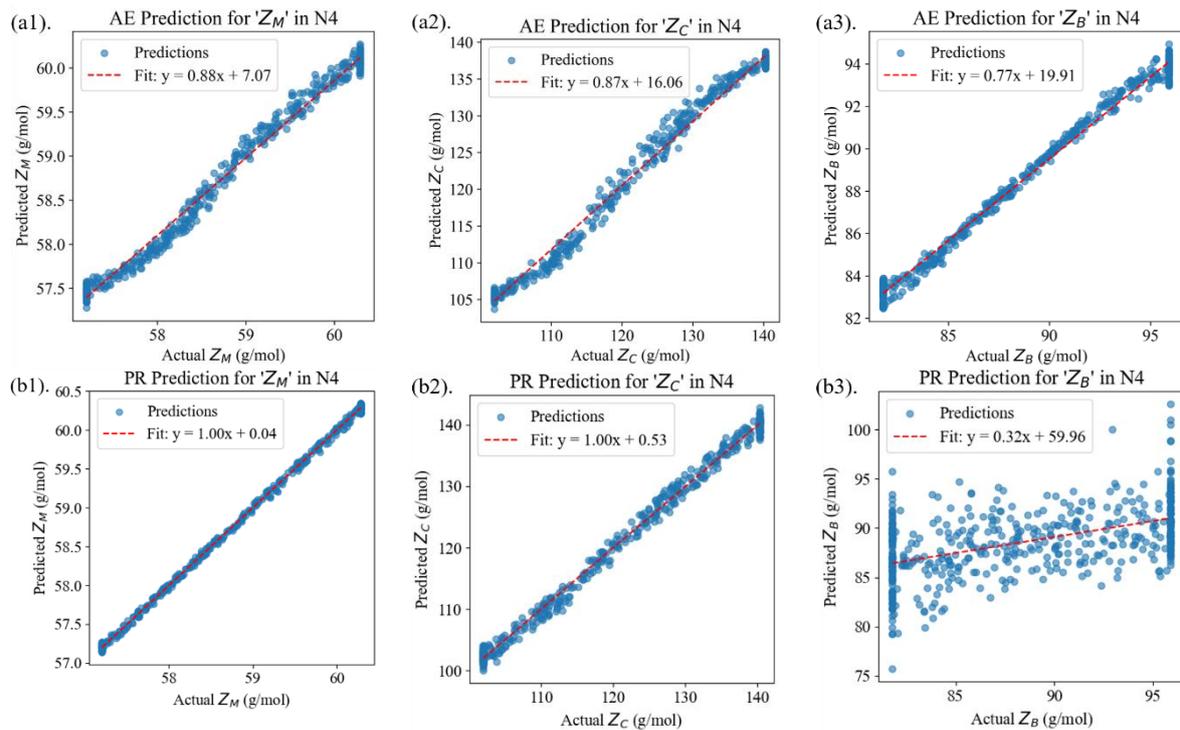

**Fig. 11.** Comparison between the predicted and actual values for $Z_M$, $Z_C$, $Z_B$: **(a1, a2, a3)**. $Z_M$, $Z_C$, $Z_B$ from AE model, **(b1, b2, b3)**. $Z_M$, $Z_C$, $Z_B$ from the PR model.

The predicted value of the molecular mass of ceramic ($Z_C$), braze alloy ($Z_B$) and metal alloy ($Z_M$) for some random input parameters/features are listed in Tables 10 and 11, respectively.

**Table 10:** Predicted $Z_C$, $Z_B$, and $Z_M$ for random input parameters/features from the AE model.

| Input features | | | | | | | Masked feature | | |
|---|---|---|---|---|---|---|---|---|---|
| Porosity (%) | VMS (MPa) | $\alpha_{eff}$ (°K$^{-1}$) | $\alpha_C$ (°K$^{-1}$) | $\alpha_B$ (°K$^{-1}$) | $\alpha_M$ (°K$^{-1}$) | $Z_{eff}$ (g/mol) | Pred. $Z_C$ (g/mol) | Pred. $Z_M$ (g/mol) | Pred. $Z_B$ (g/mol) |
| 0.166073 | 118 | .969×10$^{-5}$ | .748×10$^{-5}$ | 1.82×10$^{-5}$ | 1.16×10$^{-5}$ | 69.94 | 107.10 | 58.47 | 93.62 |
| 0.121712 | 166 | .979×10$^{-5}$ | .533×10$^{-5}$ | 1.94×10$^{-5}$ | 1.45×10$^{-5}$ | 72.96 | 123.31 | 59.80 | 87.13 |

**Table 11:** Predicted $Z_C$, $Z_B$, and $Z_M$ for random input parameters/features from the PR model.

| Input Features | | | | Output parameters and error | | | | | | | | |
|---|---|---|---|---|---|---|---|---|---|---|---|---|
| Porosity (%) | VMS (MPa) | $\alpha_{eff}$ (°K$^{-1}$) | $Z_{eff}$ (g/mol) | Act. $Z_C$ (g/mol) | Pred. $Z_C$ (g/mol) | Error (%) | $Z_M$ (g/mol) | Pred. $Z_M$ (g/mol) | Error (%) | Act. $Z_B$ (g/mol) | Pred. $Z_B$ (g/mol) | Error (%) |
| 0.083474 | 119 | 8.34×10$^{-6}$ | 71.51 | 120.40 | 121.39 | 0.01 | 57.75 | 57.74 | 0.01 | 93.40 | 89.01 | 4.7 |
| 0.300623 | 210 | 9.58×10$^{-6}$ | 75.23 | 137.21 | 136.44 | 0.562 | 60.29 | 60.33 | 0.07 | 81.79 | 86.40 | 5.64 |

The coefficient of thermal expansion material property for metal material is represented by $\alpha_M$ is also predicted by the implemented AE and PR model. The plot for comparing the predicted and actual values is shown in Fig. 12. The plot depicts that the fitting equation for the AE and PR model is $y = 0.92x$, and $y = 1.00x$ respectively. The higher accuracy of PR for the prediction feature is due to the inclusion of third-degree polynomials and interaction terms. The linear behavior shows the implemented models are accurate.

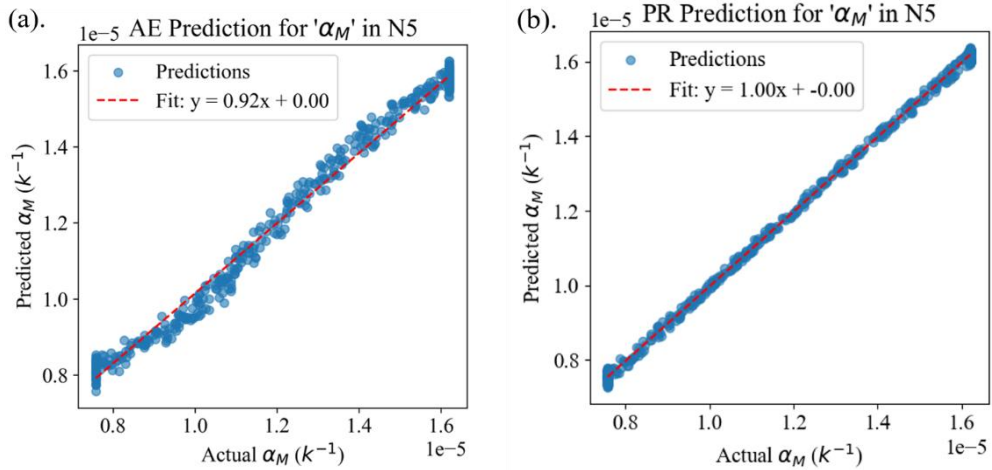

**Fig. 12.** Comparison between the predicted and actual values for $\alpha_M$ (CTE of metal): (**a**). AE model, (**b**). PR model.

The predicted value of $\alpha_M$ from the developed AE model and PR model for some given random input parameters/features are presented in Table 12 and Table 13.

**Table 12:** Predicted $\alpha_M$ for random input parameters/features from the AE model.

| Input features | | | | | | | | | Masked feature | | Error (%) |
|---|---|---|---|---|---|---|---|---|---|---|---|
| Porosity (%) | VMS (MPa) | $\alpha_{eff}$ (°K$^{-1}$) | $\alpha_C$ (°K$^{-1}$) | $\alpha_B$ (°K$^{-1}$) | $Z_{eff}$ (g/mol) | $Z_C$ (g/mol) | $Z_B$ (g/mol) | $Z_M$ (g/mol) | Act. $\alpha_M$ (°K$^{-1}$) | Pred. $\alpha_M$ (°K$^{-1}$) | |
| 0.423226 | 154 | 1.13×10$^{-5}$ | .792×10$^{-5}$ | 1.94×10$^{-5}$ | 70.15 | 102.61 | 87.15 | 60.29 | 1.62×10$^{-5}$ | 1.57×10$^{-5}$ | 3.09 |
| 0.392819 | 123 | .844×10$^{-5}$ | .474×10$^{-5}$ | 2.00×10$^{-5}$ | 72.57 | 128.44 | 82.03 | 58.10 | 1.01×10$^{-5}$ | .956×10$^{-5}$ | 5.34 |

**Table 13:** Predicted $\alpha_M$ for random input parameters/features from the PR model.

| Input parameters | | | | | | Output parameter | | Error (%) |
|---|---|---|---|---|---|---|---|---|
| Porosity (%) | $VMS$ (MPa) | $\alpha_{eff}$ (°K$^{-1}$) | $Z_{eff}$ (g/mol) | $\alpha_C$ (°K$^{-1}$) | $\alpha_B$ (°K$^{-1}$) | Actual $\alpha_M$ (°K$^{-1}$) | Predicted $\alpha_M$ (°K$^{-1}$) | |
| 0.271358 | 124 | .802×10$^{-5}$ | 73.06 | 4.25×10$^{-6}$ | 1.98×10$^{-5}$ | 9.36×10$^{-6}$ | 9.32×10$^{-6}$ | 0.43 |
| 0.064738 | 103 | .808×10$^{-5}$ | 70.71 | 5.97×10$^{-6}$ | 1.90×10$^{-5}$ | 7.58×10$^{-6}$ | 7.50×10$^{-6}$ | 1.05 |

The feature variable $\alpha_C$ (CTE of ceramic material component) of N6 configuration of input-output features is predicted, and a plot of comparison for the AE and PR model is given in 13. It is observed that the predicted values deviate slightly for the AE model, while for the PR model, the values lie near the linear fit. The best-fit equation for AE and PR models is $y = 0.92x$, and $y = 1.00x$, respectively. The linear correlation between the actual and predicted values shows the efficiency of the developed models in accurately predicting the features.

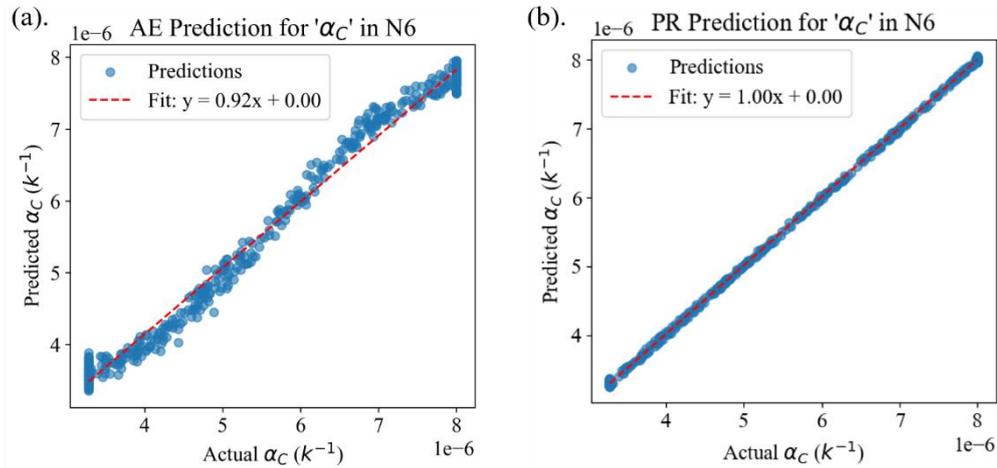

**Fig. 13.** Comparison between the predicted and actual values for $\alpha_C$ (CTE of ceramic): (**a**). AE model, (**b**). PR model.

The predicted value of $\alpha_C$ for some random input parameters from the developed AE and PR models are listed in Table 14 and Table 15 respectively.

**Table 14:** Predicted $\alpha_C$ for random input parameters/features from the AE model.

| Input features | | | | | | | | | Masked feature | | Error (%) |
|---|---|---|---|---|---|---|---|---|---|---|---|
| Porosity (%) | $VMS$ (MPa) | $\alpha_{eff}$ (°K$^{-1}$) | $\alpha_B$ (°K$^{-1}$) | $\alpha_M$ (°K$^{-1}$) | $Z_{eff}$ (g/mol) | $Z_C$ (g/mol) | $Z_B$ (g/mol) | $Z_M$ (g/mol) | Act. $\alpha_C$ (°K$^{-1}$) | Pred. $\alpha_C$ (°K$^{-1}$) | |
| 0.166073 | 118 | .969×10$^{-5}$ | 1.82×10$^{-5}$ | 1.16×10$^{-5}$ | 69.94 | 106.17 | 95.91 | 58.63 | .748×10$^{-5}$ | .758×10$^{-5}$ | 1.34 |
| 0.121712 | 166 | .980×10$^{-5}$ | 1.94×10$^{-5}$ | 1.45×10$^{-5}$ | 72.96 | 123.65 | 87.05 | 59.69 | .533×10$^{-5}$ | .535×10$^{-5}$ | 0.37 |

**Table 15:** Predicted $\alpha_C$ for random input parameters/features from the PR model.

| Input parameter | | | | | | Output parameter | | Error (%) |
|---|---|---|---|---|---|---|---|---|
| Porosity (%) | VMS (MPa) | $\alpha_{eff}$ (°K$^{-1}$) | $Z_{eff}$ (g/mol) | $\alpha_B$ (°K$^{-1}$) | $\alpha_M$ (°K$^{-1}$) | Actual $\alpha_C$ (°K$^{-1}$) | Predicted $\alpha_C$ (°K$^{-1}$) | |
| 0.124578 | 125 | 1.04×10$^{-5}$ | 69.67 | 1.83×10$^{-5}$ | 1.34×10$^{-5}$ | 8.0×10$^{-6}$ | 8.01×10$^{-6}$ | 0.125 |
| 0.363702 | 118 | 8.54×10$^{-6}$ | 71.64 | 1.96×10$^{-5}$ | .966×10$^{-5}$ | 5.49×10$^{-6}$ | 5.51×10$^{-6}$ | 0.364 |

The developed AE and PR models are tested for the prediction of $\alpha_B$ feature of N7 configuration. The plot between the actual and predicted $\alpha_B$ values from the AE and PR model are given in Fig. 14.

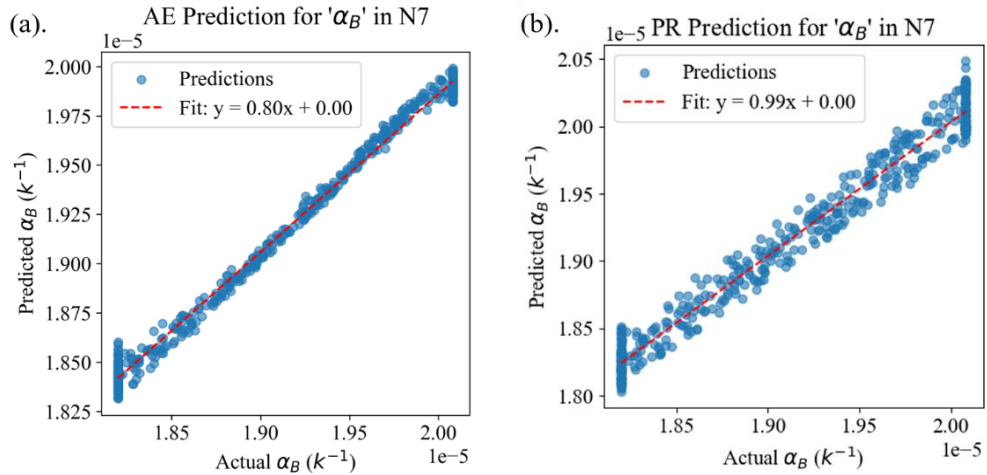

**Fig. 14.** Comparison between the predicted and actual values for $\alpha_B$ (CTE of braze): (**a**). AE model, (**b**). PR model.

The plot highlights the linear correlation between the actual and predicted values with best-fit equation $y = 0.80x$, and $y = 0.99x$ for AE and PR models, respectively. The PR model also shows a slight deviation from the linear fit due to the regression behavior of the model and also the global feature importance of $\alpha_B$ is lower. The predicted value of $\alpha_B$ from the developed AE and PR models are given in Table 16 and Table 17, respectively.

**Table 16:** Predicted $\alpha_B$ for random input parameters/features from AE model.

| Input features | | | | | | | | | Masked feature | | Error (%) |
|---|---|---|---|---|---|---|---|---|---|---|---|
| Porosity (%) | VMS (MPa) | $\alpha_{eff}$ (°K$^{-1}$) | $\alpha_C$ (°K$^{-1}$) | $\alpha_M$ (°K$^{-1}$) | $Z_{eff}$ (g/mol) | $Z_C$ (g/mol) | $Z_B$ (g/mol) | $Z_M$ (g/mol) | Act. $\alpha_B$ (°K$^{-1}$) | Pred. $\alpha_B$ (°K$^{-1}$) | |
| 0.124578 | 125 | 1.04×10$^{-5}$ | .80×10$^{-5}$ | 1.34×10$^{-5}$ | 69.67 | 101.96 | 95.38 | 59.29 | 1.83×10$^{-5}$ | 1.85×10$^{-5}$ | 1.09 |
| 0.363702 | 118 | .854×10$^{-5}$ | .549×10$^{-5}$ | .966×10$^{-5}$ | 71.64 | 122.38 | 85.19 | 57.95 | 1.96×10$^{-5}$ | 1.95×10$^{-5}$ | .510 |

Table 17: Predicted $\alpha_B$ for random input parameters/features from the PR model.

| Input parameters | | | | | | Output parameters | | Error (%) |
|---|---|---|---|---|---|---|---|---|
| Porosity (%) | VMS (MPa) | $\alpha_{eff}$ (°K$^{-1}$) | $Z_{eff}$ (g/mol) | $\alpha_C$ (°K$^{-1}$) | $\alpha_M$ (°K$^{-1}$) | Actual $\alpha_B$ (°K$^{-1}$) | Predicted $\alpha_B$ (°K$^{-1}$) | |
| 0.117541 | 196 | .958×10$^{-5}$ | 75.20 | 3.78×10$^{-6}$ | 1.62×10$^{-5}$ | 1.95×10$^{-5}$ | 1.94×10$^{-5}$ | .510 |
| 0.064738 | 103 | .808×10$^{-5}$ | 70.71 | 5.97×10$^{-6}$ | .758×10$^{-5}$ | 1.90×10$^{-5}$ | 1.93×10$^{-5}$ | 1.58 |

The multi-output prediction performance of both the AE and PR models is also tested for CTE feature variables (namely. $\alpha_M, \alpha_C,$ and $\alpha_B$ ) of N8 configuration. The plot for all the feature variables is given in Fig. 15. The best-fit equation for the AE and PR model is $y = 0.88x, y = 0.87x, y = 0.78x$ and $y = 1.00x, y = 1.00x, y = 0.32x$ respectively for $\alpha_M, \alpha_C,$ and $\alpha_B$ features.

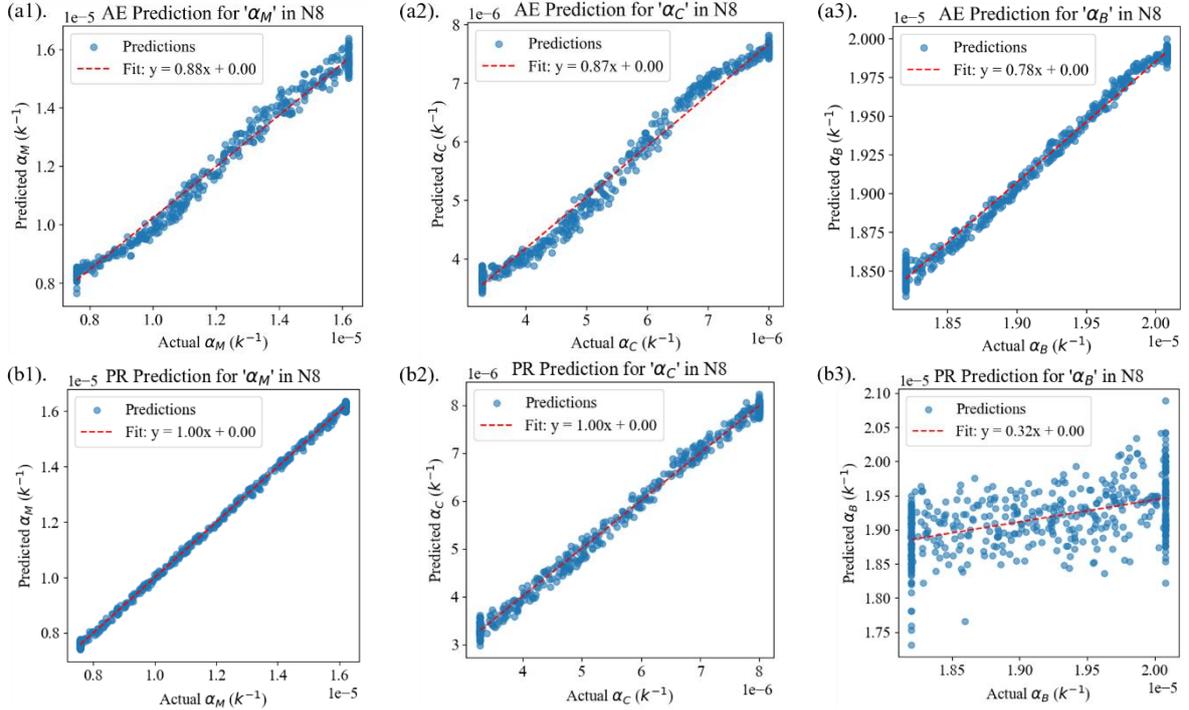

**Fig. 15.** Comparison between the predicted and actual values for $\alpha_M, \alpha_C, \alpha_B$: (**a1, a2, a3**). $\alpha_M, \alpha_C, \alpha_B$ from AE model, (**b1, b2, b3**). $\alpha_M, \alpha_C, \alpha_B$ from the PR model.

It is observed that the AE model accurately predicts all the features with a slight deviation from the linear fit, as shown in Fig. 15(a1), 15(a2), and 15(a3). The PR model accurately predicts for $\alpha_M$ and $\alpha_C$ parameters as shown in Fig. 15(b1) and 15(b2) but a high deviation is observed for $\alpha_B$ (CTE of braze material) as shown in Fig. 15(b3). A correlation of $y = 0.32x$ is established from the PR model. This may be due to the lower feature importance value of $\alpha_B$ parameter. Hence, the results show that the autoencoder (AE) model is suitable for single- and multi-output feature prediction, while PR is accurate and efficient for single-output feature prediction. Also, the developed AE model can accurately predict features with lower global importance, such as $\alpha_B$ and $Z_B$ where PR is not so efficient.

The predicted CTE values of ceramic ($\alpha_C$), braze alloy ($\alpha_B$) and metal alloy ($\alpha_M$) for some random input parameters from the developed AE and PR models are given in Tables 18 and 19, respectively.

**Table 18:** Predicted $\alpha_C$, $\alpha_B$, and $\alpha_M$ for random input parameters/features from the AE model.

| Input features | | | | | | | Masked feature | | |
|---|---|---|---|---|---|---|---|---|---|
| Porosity (%) | $VMS$ (MPa) | $\alpha_{eff}$ (°K⁻¹) | $Z_C$ (g/mol) | $Z_B$ (g/mol) | $Z_M$ (g/mol) | $Z_{eff}$ (g/mol) | Pred. $\alpha_C$ (°K⁻¹) | Pred. $\alpha_M$ (°K⁻¹) | Pred. $\alpha_B$ (°K⁻¹) |
| 0.14932 | 209 | .942×10⁻⁵ | 137.48 | 94.54 | 60.29 | 75.64 | .376×10⁻⁵ | 1.59×10⁻⁵ | 1.84×10⁻⁵ |
| 0.280423 | 90.6 | .889×10⁻⁵ | 101.96 | 83.39 | 57.2 | 68.04 | .778×10⁻⁵ | .822×10⁻⁵ | 1.99×10⁻⁵ |

**Table 19:** Predicted $\alpha_C$, $\alpha_B$, and $\alpha_M$ for random input parameters/features from the PR model.

| Input parameters | | | | Output parameters and error | | | | | | | | |
|---|---|---|---|---|---|---|---|---|---|---|---|---|
| Porosity (%) | $VMS$ (MPa) | $\alpha_{eff}$ (°K⁻¹) | $Z_{eff}$ (g/mol) | Act. $\alpha_C$ (°K⁻¹) | Pred. $\alpha_C$ (°K⁻¹) | Error (%) | $\alpha_M$ (°K⁻¹) | Pred. $\alpha_M$ (°K⁻¹) | Error (%) | Act. $\alpha_B$ (°K⁻¹) | Pred. $\alpha_B$ (°K⁻¹) | Error (%) |
| 0.46995 | 119 | 8.58×10⁻⁶ | 71.92 | 5.23×10⁻⁶ | 5.28×10⁻⁶ | 0.96 | 9.94×10⁻⁶ | 9.91×10⁻⁶ | 0.30 | 2.01×10⁻⁵ | 1.99×10⁻⁵ | 0.99 |
| 0.287374 | 126 | 8.43×10⁻⁶ | 71.87 | 5.37×10⁻⁶ | 5.46×10⁻⁶ | 1.68 | 9.70×10⁻⁶ | 9.68×10⁻⁶ | 0.21 | 1.92×10⁻⁵ | 1.89×10⁻⁵ | 1.56 |

The developed models are also tested for the strength prediction (defined by $VMS$ feature/parameter) of the joint assembly as a function of material property as input features. The plot of $VMS$ prediction from the AE and PR models is plotted and shown in Fig. 16.

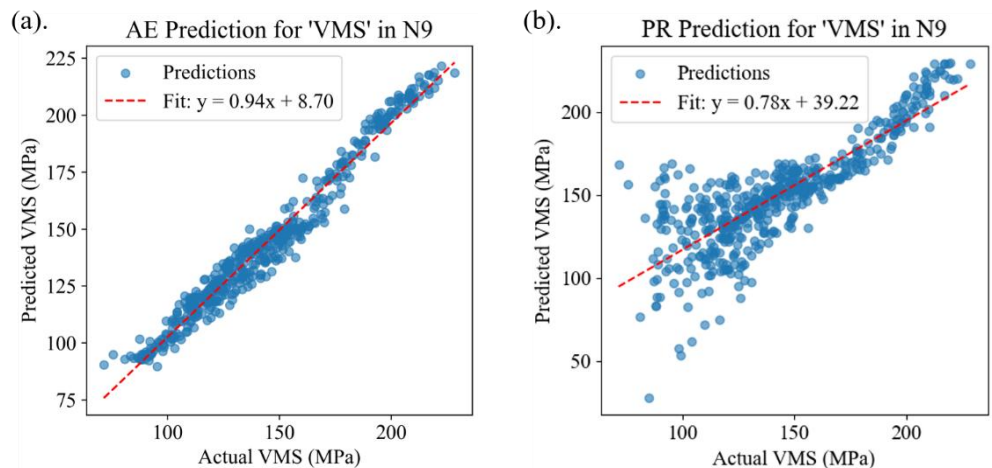

**Fig. 16.** Comparison between the predicted and actual values for $VMS$: (**a**). AE model, (**b**). PR model.

The plot shows that the fitting equation from the AE and PR model is $y = 0.94x + 8.70$ and $y = 0.78x+39.22$, respectively. That is, the performance of the AE model for $VMS$ prediction is higher than PR. Also, the plot is more scattered and deviates from the linear fit for PR, while the plot for the AE model converges and differs slightly from the linear fit. That means the AE model is more robust and adaptive due to encoder-decoder architecture than the PR model. Hence, the developed model can predict any of the material properties of brazed ceramic-metal composite material as a function of strength/quality and vice versa. The predicted value of $VMS$ from the developed AE and PR model for some random input parameters is listed in Table 20 and Table 21.

**Table 20:** Predicted $\alpha_B$ for random input parameters/features from the AE model.

| Input Parameters | | | | | | | | | Output parameter | | Error (%) |
|---|---|---|---|---|---|---|---|---|---|---|---|
| Porosity (%) | $\alpha_{eff}$ (°K$^{-1}$) | $Z_{eff}$ (g/mol) | $Z_M$ (g/mol) | $Z_C$ (g/mol) | $Z_B$ (g/mol) | $\alpha_C$ (°K$^{-1}$) | $\alpha_M$ (°K$^{-1}$) | $\alpha_B$ (°K$^{-1}$) | Act. VMS (MPa) | Pred. VMS (MPa) | |
| 0.087434 | .769×10$^{-5}$ | 72.05 | 57.2 | 128.26 | 84.40 | 4.76×10$^{-6}$ | .758×10$^{-5}$ | 1.97×10$^{-5}$ | 112 | 118 | 5.36 |
| 0.290112 | .808×10$^{-5}$ | 74.56 | 58.44 | 140.31 | 81.79 | 3.28×10$^{-6}$ | 1.10×10$^{-5}$ | 2.01×10$^{-5}$ | 157 | 140 | 10.82 |

**Table 21:** Predicted $\alpha_B$ for random input parameters/features from the PR model.

| Input Parameters | | | | | | | | | Output parameter | | Error (%) |
|---|---|---|---|---|---|---|---|---|---|---|---|
| Porosity (%) | $\alpha_{eff}$ (°K$^{-1}$) | $Z_{eff}$ (g/mol) | $Z_M$ (g/mol) | $Z_C$ (g/mol) | $Z_B$ (g/mol) | $\alpha_C$ (°K$^{-1}$) | $\alpha_M$ (°K$^{-1}$) | $\alpha_B$ (°K$^{-1}$) | Act. VMS (MPa) | Pred. VMS (MPa) | |
| 0.25096 | 1.1×10$^{-5}$ | 70.14 | 59.97 | 104.41 | 84.16 | 7.7×10$^{-6}$ | 1.53×10$^{-5}$ | 1.98×10$^{-5}$ | 149 | 160 | 7.38 |
| 0.161211 | .790×10$^{-5}$ | 74.67 | 58.27 | 140.31 | 86.84 | 3.28×10$^{-6}$ | 1.06×10$^{-5}$ | 1.94×10$^{-5}$ | 151 | 165 | 9.27 |

## 4. Discussion

The features/parameters used in training the AI models include the material property, namely CTE and molecular mass, strength/quality assessment parameter, which is VMS, and micro-structure property, namely Porosity of the joint assembly. The Von Mises Stress values obtained from the FEM simulation are used in training the different AI models. This is due to the availability of the FEM simulation data from our previous publication [9]. The structural parameter data obtained from NDT techniques such as X-ray CT (porosity or void fraction) or DT techniques such as compression test (tensile or ultimate strength) may also be utilized in the training and development of the model. Also, we have used molecular mass and CTE parameter data of materials due to availability in the reported literature. Other parameters, such as specific heat, density, conductivity, etc., can also be used. The current methodology has not explicitly incorporated noise injection or data augmentation from real-world experiment data; however, the AE model—offers potential robustness due to its latent feature compression. The motivation of this study is to propose a methodology to predict the materials for brazed ceramic-metal joint assemblies. The methodology for inclusiveness of other techniques or material properties will remain the same; however, the results may vary depending upon the deciding parameters used in the analysis.

The different ML/DL models are trained and tested: LR, PR, RF, ANN, and one multi-output Autoencoder (AE) model. The developed PR model outperforms the prediction of single features having high feature importance. However, AE models outperform the prediction of multi-output variables simultaneously and accurately predict the single-output features with low feature importance. Hence, the developed AE model can predict multiple features. The model can capture the global as well as local features importance. Also, unlike other ML/DL models, the separate training of the AE model for different input-output feature combinations is not required. The model is trained once with an equal number of input-output features. The trained AE model can then predict the features according to the requirement. The AE model developed in this study predicts the material property of brazed ceramic-metal composite materials as a function of strength parameter (VMS) and Porosity (micro-structure property). The vice-versa also holds.

The optimal material/combination of materials that is the one possessing lower coefficient of thermal expansion and molecular mass, which yields lower Von Mises/thermal stress, can also be selected. For example, the alumina (101.96 g/mol) ceramic, Ag-Cu-Ti (81.79 g/mol) braze alloy, and Kovar material with a molecular mass of 57.20 g/mol yields lower average VMS (87.08 MPa), as compared to alumina/Ag-Cu-Ti/Monel-400 (60.29g/mol) which yields higher VMS (147 MPa). The developed AE model predicts a molecular mass of 59.87 g/mol (~4.45% error) for Kovar material, given the molecular mass of alumina and braze alloy and the VMS values. This is because the Kovar material, due to its lower molecular mass, possesses higher stiffness and contributes to lower plastic deformation, leading to uniform stress distribution and lower average VMS. The Monel-400 metal material with higher ductility and lower stiffness may accumulate higher stress in the joint interface regions. Thus, the joint assemblies containing Alumina/Ag-Cu-Ti/Kovar materials may have higher bonding strength and larger fracture limit under given loading conditions. This matches with our previously reported study [9]. The vice-versa also holds; the material with lower molecular mass and CTE has lower VMS than higher CTE materials, as listed in Tables 20 and 21. Thus, the proposed study also highlights the selection of AI-optimized brazed ceramic-metal composite materials.

## 5. Conclusion

The proposed study provides an AI-driven methodology to predict the materials and combination of materials used for brazed ceramic-metal composite material joint assemblies. The methodology can also be used to predict the strength/quality assessment parameter of brazed ceramic-metal composite materials joint assembly as a function of material property. Four different ML/DL and one multi-output AE model are trained and tested for nine input-output feature configurations. The AE model, also referred to as the universal model, outperforms all the AI models and predicts the molecular mass and CTE parameters of different materials with an average error of ~0.16-1.26 % and ~0.55-3.78% of literature-reported values, respectively. The strength parameter, namely VMS, is also tested for prediction from the developed AE model, and the predicted lies with an average error of ~3.61% of the values obtained from the simulation. A small error between the material property reported in the literature and the developed model depicts the accuracy of the proposed AI-driven methodology. The methodology can be used to predict the new materials and combinations of materials that can be used in fabricating brazed-ceramic metal composite material joint assemblies.

Future Work: Future studies will explore training with controlled noise perturbations, including the real-world data obtained from experiments performed on brazed ceramic-metal joint assemblies and testing more optimal material combinations.

Acknowledgments: SK is thankful to the IITR Institute Assistantship.

Author contributions: SK: Writing, Methodology, Data-analysis, and Code Verification; MG: Funding, writing, and Methodology; VK: Code writing and verification: RKV: Funding.

Funding: STC-1937-PHY by ISRO.

Data and material availability: Data will be provided on request.

Declarations

Conflict of interest: The authors have applied for a patent at IPO including the content of this work.

Ethics approval and consent to participate: Not required/applicable.

Consent for publication: All authors have read the manuscript and provided their consent after significant contribution.